\documentclass[a4paper,nofootinbib,twocolumn]{revtex4}
\usepackage{amsmath,amsfonts,amssymb}
\usepackage{hyperref}
\usepackage{mathrsfs}
\usepackage{graphicx}

\newcommand{\ie}{\emph{i.e.}~}

\newcommand{\ket}[1]{\left| #1 \right \rangle}
\newcommand{\bra}[1]{\left\langle #1 \right |}
\newcommand{\braket}[2]{\left\langle #1 | #2 \right\rangle}
\newcommand{\ha}{\hat{a}}
\newcommand{\hrho}{\hat{\rho}}
\newcommand{\had}{\hat{a}^{\dagger}}

\newcommand{\hphi}{\hat{\varphi}}
\newcommand{\hphid}{\hat{\varphi}^{\dagger}}

\newcommand{\hxi}{\hat{\xi}}
\newcommand{\hxid}{{\hat{\xi}^{\dagger}}}

\newcommand{\Tr}[1]{\mathrm{Tr}\left(#1 \right)}

\newcommand{\fv}{\left| \Omega \right\rangle}
\newcommand{\fvbra}{\left\langle \Omega \right|}

\newcommand{\be}{\nopagebreak[3]\begin{equation}}
\newcommand{\ee}{\end{equation}}
\newcommand{\bee}{\nopagebreak[3]\begin{equation*}}
\newcommand{\eee}{\end{equation*}}
\newcommand{\ba}{\nopagebreak[3]\begin{eqnarray}}
\newcommand{\ea}{\end{eqnarray}}
\newcommand{\baa}{\nopagebreak[3]\begin{eqnarray*}}
\newcommand{\eaa}{\end{eqnarray*}}

\newcommand{\id}{\mathbb{I}}

\newcommand{\iu}{\mathrm{i}}
\newcommand{\CC}{\hat{\mathcal{C}}}
\newcommand{\DD}{{\mathcal{D}}}
\newcommand{\OO}{{\mathcal{O}}}
\newcommand{\OOO}{\hat{\mathcal{O}}}
\newcommand{\OOop}{{\mathrm{O}}}
\newcommand{\Aop}{\hat{\mathrm{A}}}
\newcommand{\Aexp}{{\mathcal{A}}}
\newcommand{\eol}{\nonumber\\&}

\newcommand{\WW}{\mathbb{W}}
\newcommand{\PP}{\mathbb{P}}
\newcommand{\QQ}{\mathbb{Q}}

\newcommand{\VV}{\mathcal{V}}
\newcommand{\KV}{\mathscr{V}}

\newcommand{\SU}{\mathrm{SU}(2)}

\newcommand{\Prob}{\mathcal{P}}

\newcommand{\eff}{{\mathrm{eff}}}

\begin{document}

\title{Effective equations for GFT condensates from fidelity}
\author{Lorenzo Sindoni}
\email{sindoni@aei.mpg.de}
\affiliation{Max Planck Institute for Gravitational Physics \\
(Albert Einstein Institute) \\
Am M¬uhlenberg 1, 14476 Golm, Germany, EU}
\begin{abstract}
The derivation of effective equations for group field theories is discussed from a variational
point of view, with the action being determined by the fidelity of the trial state with respect
to the exact state.
It is shown how the maximisation procedure with respect to the parameters of the trial state
lead to the expected equations, in the case of simple condensates. Furthermore, we show that 
the second functional derivative of the fidelity gives a compact way to estimate, within the effective
theory itself, the limits of its validity. The generalisation can be extended to include
the Nakajima--Zwanzig projection method for general mixed trial states.
\end{abstract}

\maketitle

\section{Introduction}

A general procedure to derive effective equations for (a modified form of) quantum Cosmology
has been described in \cite{cosmoshort,cosmolong}. The proposal hinges on the identification, within Group Field Theory (GFT) models for quantum gravity, of certain families of states, GFT condensates, that naturally identify homogeneous (but possibly 
anisotropic) spacetimes.\footnote{These states are essentially the GFT analogue of coherent states (and extensions as squeezed states).}
The imposition of the equations of motion of the theory (in a sense that will be discussed later in
detail) results into the 
effective equations for the order parameter controlling the quantum states. 
This result opens a possible path to the derivation of effective equations for (quantum) cosmology
directly from a microscopic
model \cite{Gielen,GielenOriti,Calcagni}. 

Several questions are still unanswered. The objective of this paper is to examine a
specific problem which has not been considered in detail: how good are 
condensed states (and the effective description for cosmology that they imply) as an approximation for the exact states of the full theory?
Indeed, one of the crucial aspects for the assessment of the validity of the results of an
effective theory is the estimate of the error introduced by the particular truncation that
has been used to derive it. As the models considered are not defined in terms of a geometrical
background, traditional approaches \cite{Burgess}, based on dimensional analysis (determining the terms to be included in the effective theory, given the desired accuracy), are of no
immediate use. 

The goal of this paper is then twofold. On one hand, we establish a criterion to determine, 
\emph{within the effective theory}, its own limits of validity. Second, with the particular method considered, we
describe a general method that allows the use of tools of (quantum) statistical mechanics
to a special class of timeless systems.

In order to do so, we need to reconsider the interplay between the trial state and the equations
of motion. In previous work the dynamics was imposed onto the trial state examining the
Schwinger--Dyson hierarchy and imposing the validity of some of the equations at least
on average. While correct, this complicates the estimate of the validity of the resulting effective theory, as it is difficult to say how to truncate the hierarchy. A more systematic method is needed.

In order to achieve these results we will use the notion of \emph{fidelity}
\cite{fidelity}, the projection of the trial state (which we wish to use to develop the effective theory)
onto the physical state. The maximisation of the fidelity is then used to derive an effective theory for the GFT condensates that follow most closely
the physical state.
As a bonus, we will establish a precise connection with methods and concepts of nonequilibrium
statistical mechanics, and in particular with the Nakajima--Zwanzig formalism \cite{openquantumsystems,Balian}, which can be used, in principle, to deal with trial states modeled directly in terms of the physical observables that
we are interested in, instead of passing through ``condensate wavefunctions'' as in \cite{cosmoshort,cosmolong}.

Using a specific class of models which can be treated in great detail, as the formal expression
of the exact state is known, there are three outcomes. First, the effective equations
are now derived from an information-theoretic perspective, in a systematic way: their derivation 
uses a variational principle on the overlap between the microscopic
state and the representative macroscopic state that stores the information about a
restricted set of operators that we can choose as physically relevant. 
Effective actions for composite operators, in such calculations, assume a precise
meaning in determining the expectation values and the properties of the effective
theory.
Second, it is now possible to concretely assess the validity of the effective equations
for cosmology, by checking the peakedness of the fidelity with respect to the variables
of interest. 
The effects of the approximation related to the truncation of the state, and going beyond
the conditions required merely by self-consistency, can now be estimated, we stress it again, {within
the effective theory itself}.
Third, a precise kind of effects is expected, namely the presence of statistical
fluctuation around the mean field, that can be quantitatively estimated. Their origin is purely quantum
gravitational, as opposed to the usual situations in which they encode the
fluctuations in the stress energy tensor of matter fields. These results then extends
stochastic gravity to include quantum gravity fluctuations. 

The paper is organised as follows. In section II a brief review of GFTs is given.
In section III a very simple case is considered and described in details. The
generalisation to a larger class of (mixed) states is described in section IV, while section V contains
a description of the application of the method to second quantised observables.
Finally, section VI contains a summary of the paper, while in section VII some final
remarks about future developments are presented. 

\section{Brief review}

Group Field Theories (GFTs) are particular quantum field theories on group manifolds, 
which can be designed to define a proper sum over geometries quantisation of the gravitational field. They are built
generalising the techniques used for matrix models for 2D gravity \cite{matrixmodels}
and tensor models \cite{GurauRyan}: their perturbative expansion around the Gaussian theory generates Feynman diagrams that can be interpreted in terms of discrete geometries
of the chosen dimension. Besides the topology, encoded in the combinatorics of the Feynman
graphs, the theory contains additional data, interpreted in terms of metric and connection variables, suitably discretised. The construction for quantum gravity models is then completed once the amplitudes are determined by the chosen discretisation of a gravitational action \cite{OritiMicroDyn}.

To fix the ideas, we consider for the moment the simplest version of the four dimensional theory,
in which the field operators are defined over four copies of SU(2), which is
the relevant group for most of the models considered in the literature. The field operators $\hphi(g_1,g_2,g_3,g_4)$ and $\hphid(g_1,g_2,g_3,g_4)$, 
possess a special gauge invariance property, under the diagonal action of $SU(2)$
on $SU(2)^4$:
\begin{equation}
\hphi(g_{1}h,g_{2}h,g_{3}h,g_{4}h) = 
\hphi(g_{1},g_{2},g_{3},g_{4})
\end{equation}
and they obey the bosonic commutation relations 
\begin{equation}[\hphi(g_1,g_2,g_3,g_4),\hphid(h_1,h_2,h_3,h_4)] = \prod_{i=1}^{4}\int dk \delta(g_i k h_{i}^{-1})\,.
\end{equation}
In the following a more condensed notation will sometimes be needed, when
no confusion arises, in which a quartet of group
elements like $g_{1},g_{2},g_{3},g_{4}$ will be denoted simply by $g_{I}$.

The ladder operators built in this way can be interpreted as operators creating or destroying
a very special kind of ``particle'', interpreted as
4-valent vertices, dual to tetrahedra, and carrying geometrical data associated to
the group elements $g_I$. States in the Fock space, then, can be interpreted as ensembles of  tetrahedra treated as identical particles. The wavefunction can be ultimately be used to consider states
in which these tetrahedra are glued together along common faces.

This geometric interpretation in terms of tetrahedra has a clear motivation in the so-called noncommutative representation \cite{Baratin,GFTnoncomm},
obtained looking at the noncommutative Fourier transform of the field operators on the Lie algebra
$\mathfrak{su}(2)^4$:
\begin{equation}
\hat{\tilde{\phi}}(B_{1},B_{2},B_{3},B_{4}) \equiv \int (dg)^4 \prod_{i=1}^{4}e_{g_{i}}(B_{i})
\hphi(g_{1},g_{2},g_{3},g_{4}),
\end{equation}
where $e_{g_i}(B_i)$ are noncommutative plane waves.
In this version, the ladder operators can be directly interpreted as operators associated to the
creation or annihilation of tetrahedra whose bivectors (the wedge product of pairs of vectors associated to each edge of the tetrahedron) are given by $B_{1},B_{2},B_{3},B_{4}$, with
the gauge invariance condition being translated into the closure condition $B_{1}+B_{2}+B_{3}+B_4=0$. Furthermore, the perturbative expansion leads naturally to an interpretation
of the amplitudes at fixed graph in terms of simplicial gravity path integrals on a fixed triangulation,
as a direct inspection of the amplitudes for gravitational models confirms.

A generic state with $n$ particles, in the Fock space, can be built in terms of creation operators
acting on the Fock space:
\begin{equation}
\ket{\psi} = \int (dg)^{4n} \psi(g_{I_1},\ldots,g_{I_n}) 
\hphid(g_{I_1})
\ldots
\hphid(g_{I_n})
\fv ,
\end{equation}
where $\fv$ denotes the Fock vacuum, $\hphi(g_{I})\fv = 0$, and $\psi$ is a $n-$body wavefunction. Notice that this latter is completely symmetric under permutation of its arguments,
as the field operators obey the bosonic statistics.
It turns out that certain specific shapes of $\psi$ (associated to a specific pattern of
convolutions between arguments of the wavefunction, as dictated by an underlying graph) can be  shown to coincide with the spin network wavefunctions of Loop Quantum Gravity (LQG) \cite{2nd}.

The dynamics of the theory, in general, is determined by a quantum equation of motion:
\begin{equation}
\left(\int (dh)^{4} \mathcal{K}(g_I,h_{I}) \hphi(h_{I}) + \frac{\delta \mathcal{V}}{\delta \overline{\phi(g_{I})}}[\hphi,\hphid] \right) \ket{\Psi}= 0
\label{GFTEOM}
\end{equation}
where $\mathcal{K}$ is a kinetic term kernel encoding the free propagation of the field, and
$\mathcal{V}$ is a piece of an action functional that encodes the interacting part of the theory.
It contains, generically, (nonlocal) monomials in the field operators of degree higher than two. 
This implies that solutions of \eqref{GFTEOM} are not eigenstates of
the number operator and can contain, in principle, an arbitrary number of particles.

A perturbative solution can be formally defined starting from the Fock vacuum as the solution
of the free theory (we assume that $\mathcal{K}$ is nondegenerate).
For certain shapes of interaction term \cite{Ooguri}, the Feynman rules determined by \eqref{GFTEOM} lead
to the sum over spinfoams \cite{Perez}. In this way the
sum-over-geometries approach to quantum gravity can be seen to be contained also in this
particular operatorial formulation.

The advantage of the Fock structure is that it is easy to work with
arbitrary superpositions of states containing an arbitrary
number of ``particles'', as the solutions to \eqref{GFTEOM}. At the same time, it gives
the possibility of dealing with a completely new class of trial states to be used as
approximate solutions, necessary to extract concrete physical predictions.

Coherent states, squeezed states and generalisations were then exploited in \cite{cosmoshort,cosmolong,Gielen}) to derive effective equations for states that admit a natural interpretation in terms
of homogeneous cosmologies. Remarkably, this was accomplished while partially preserving the sum-over-geometries character of
the theory, as no preferred triangulation (thus identifiable as a background) is singled out.

One important aspect has not been considered explicitly. While it is possible to
elaborate a number of self-consistency conditions on the wavefunctions of the condensed states which
are to be satisfied, for the geometric interpretation to make sense, the validity of the approximation
already involved in the choice of the condensate states has not been addressed. What can be noticed is that the condensed state approximation (coherent state, dipole, etc.)
corresponds, implicitly, to the assumption that the physical state is characterized by the fact
that the $n-$point Green's functions associated to it are factorized in terms of the $1-$ or
$2-$point functions. This analysis might not be the most efficient way to estimate the error associated to the approximation,
as it is difficult to say what is the relevant set of Green's functions to be considered, and which equations of the Schwinger--Dyson hierarchy should be kept.

In this paper it will be shown that this estimate can be given in a more transparent way
using the concept of \emph{fidelity} \cite{fidelity}, \ie imposing the equation of motions in an
indirect way by asking that the overlap between the exact state of the system and the
approximate state considered is in fact maximised. While not free from ambiguities,
this prescription allows for a transparent connection with methods used in quantum statistical
mechanics and to the possibility of dealing with much more general situations than
simple condensates. 

\section{A simple case}

Let us consider the case
of the (unnormalized) states that have been introduced in \cite{cosmolong} as exact solutions
of the equations of motion for a special class of models. 
While the generality of the structure
of the states is certainly questionable (group field theories of interest for realistic models
might not offer such a neat representation of the states solving the quantum equations
of motion), they offer a very simple setting in which the general procedure can be
tested with detailed calculations, and shown to reproduce known results.

The states that we are considering in the following are (formally) exact solutions
of a group field theory designed after the Ooguri model \cite{Ooguri}, with the 
equations of motion for the physical states (to be used for four dimensional theories) of the form:
\begin{widetext}
\begin{equation}
\left( \hphi(g_I) + \lambda \int (dg) \KV(g_{I}, g_{I}',g_{I}'',g_{I}''',g_{I}'''')\times 
\nonumber\\ \hphid(g_{I}')
\hphid(g_{I}'')
\hphid(g_{I}''')
\hphid(g_{I}'''')
 \right)\ket{\Phi} = 0
 \label{specialGFT}
\end{equation}
\end{widetext}
where $\KV$ is a kernel containing the information about the dynamics of the model. For instance,
if we were considering a field theory modeled after spinfoams, 
the kernel $\KV$ would contain the details about the imposition of the simplicity constraints
that turn BF amplitudes into gravity amplitudes. The ordinary perturbative expansion in $\lambda$, around the Fock 
vacuum, can be then interpreted in
terms of a sum over discrete geometries. The quintic\footnote{The interaction term comes from a quintic monomial in the action.} nature of the kernel $\KV$ assumes that this is done in terms of simplicial complexes.

\subsection{The state}

The GFT vacuum solving \eqref{specialGFT} is easily written as:
\begin{equation}
 \ket{\Phi} = \exp\left(-\lambda \VV(\hphid)\right)\ket{0}\, ,
\end{equation}
where we are denoting
\begin{align}
\VV(\hphid) &=  \frac{1}{5}\int (dg)^{20} \KV(g_{I}, g_{I}',g_{I}'',g_{I}''',g_{I}'''') \times \nonumber
\\ &~~~~~~~~~
\times 
\hphid(g_{I})
\hphid(g_{I}')
\hphid(g_{I}'')
\hphid(g_{I}''')
\hphid(g_{I}'''')\, .
\end{align}
The norm of the state is
\begin{equation}
 Z[\Phi] = \braket{\Phi}{\Phi} = \bra{0}\exp\left(-\lambda^{*}\VV^*(\hphi)\right)
 \exp\left(-\lambda \VV(\hphid)\right)\ket{0}  . 
 \label{eq:firstpartition}
\end{equation}
Inserting a basis of coherent states for the field operators into \eqref{eq:firstpartition} one obtains:
\begin{align}
 Z[\Phi]=\frac{1}{Z_0}\bra{\Phi}\left(
 \int \DD \psi\DD\overline{\psi}
 \ket{\psi}\bra{\psi}\right)
\ket{\Phi} = \frac{Z_{GFT}}{Z_0} 
\end{align}
where $Z_0$ is the Gaussian partition function, needed for a correct normalisation, and
$Z_{GFT}$ is the ordinary vacuum partition function of the GFT model, when written in the functional
language.

At this stage, we are not assuming that this
state is normalisable, in the Fock space. It is possible to show with a simple example that it
it reasonable to expect that this state is in fact not normalisable in the Fock space inner product. Take the single oscillator version:
\begin{equation}
\ket{\psi} = \exp \left(-\frac{\lambda}{5} (\had)^5\right) \ket{0},
\end{equation}
with $\lambda$ real,
and compute the square of the norm inserting a basis of coherent states:
\begin{align}
\braket{\psi}{\psi} = \bra{0}
\exp \left(-\frac{\lambda}{5} (\ha)^5\right)
 \exp \left(-\frac{\lambda}{5} (\had)^5\right) \ket{0} = \nonumber \\ =\int \frac{dzdz^*}{\pi}
 \exp\left(- |z|^2 - \frac{\lambda}{5}z^5 - \frac{\lambda}{5}(z^*)^5\right).
\end{align}
Consider now $z=\rho e^{\iu \theta}$, and integrate in $\theta$, to obtain:
\begin{equation}
||\ket{\psi}||^2 = \int_{0}^{\infty} \frac{\rho}{\pi} \exp(-\rho)  2\pi I_{0}\left(\frac{\lambda}{5}\rho^5\right)
\end{equation}
where $I_{0}$ is a modified Bessel function. Its asymptotic behavior for
large values of its argument is
\begin{equation}
I_{0}(x) \sim\frac{e^{x}}{\sqrt{2\pi x}}\left( 1 + O(1/x) \right),
\end{equation}
and hence the norm diverges. 

This is not a real problem, in light of the maximisation
procedure that has been presented so far. In fact, an similar situation would be to find an optimal
Gaussian that maximises the overlap with an eigenstate of the position operator. 
Depending on the set of variational parameters that we consider (the position of the peak
or the position of the peak \emph{and} the spread of the Gaussian) we can obtain a finite norm
state that best approximates the position eigenstate. 
This is not trivial, however, and it requires a proper renormalisation procedure to be
put in place. It represents an additional sanity check for the method,
a further test to be passed.

\subsection{Simple condensate ansatz}
We now examine the possibility to use a condensate state modeled after a coherent state, instead of the exact state,
to describe the system. The case of dipoles is
significantly more involved and is discussed in the appendix. As explained in details in \cite{cosmoshort,cosmolong,Gielen},
the choice of condensed states reflects the assumption that the physical state is a state
in which each building block of our spacetime is characterized by the same distribution
of geometric data. 

The downside of this approach is that the condensed state
has a less direct connection to the physical observables than one we might desire: the condensate wavefunction
can be used as a sort of Wheeler--de Witt wavefunction for the geometry of a (quantum)
tetrahedron \cite{Barbieri}, with which one can finally compute the expectation values of geometric observables. We will examine the construction for states adapted directly to a set of observables
in the next sections.

The fidelity is defined as the overlap between the
physical state and our trial state, 
\begin{equation}
F^{2}[\sigma] := |\braket{\sigma}{\Phi}|^2
\end{equation}
where
\begin{equation}
\ket{\sigma} := \exp\left(-\frac{1}{2} \int dg_I |\sigma(g_I)|^2 +\int dg_I
\sigma(g)I)\hphid(g_I)
\right)\ket{0}
\end{equation}
is a (normalised) single field coherent state. The normalisation is required to make the procedure
consistent. One can imagine that the fidelity in these equations is in fact obtained as a limit
of the perturbative series. At any finite order in perturbation theory, one would be then comparing
perfectly normalisable states. Not asking that the trial state is normalised would simply
lead to spurious solutions in which the fidelity is maximised not because of the angle between
the two vectors in the Hilbert space is small, but because the trial vector has a very large norm.
The norm of the exact state, instead, is not really essential, as it is merely a multiplicative factor which is unaffected by the variational procedure.

As an immediate consequence of the choice of a coherent trial state,
\begin{align}
|\braket{\sigma}{\Phi}|^2 =|Z_{GFT}|^{-2} \exp\left( -S_{cl}[\sigma] \right),\\
S_{cl}[\sigma] = \int dg_I |\sigma(g_I)|^2 + \lambda^* V^*[\sigma^*]
+ \lambda V[\sigma],
\end{align}
where $S_{cl}$ is the classical action appearing in the path integral formulation of the
theory.

Therefore, asking that the state maximises the fidelity amounts to two conditions.
The first is that $\sigma$ has to be an extremum of the classical action: 
\begin{equation}
\frac{\delta S_{cl}}{\delta \sigma} = 0. \label{EOM1}
\end{equation}
This
confirms the expectations: we recover the Gross--Pitaevski equation of \cite{cosmoshort,cosmolong}. The second and more significant condition is that, in order for the fidelity to be maximised,
the second functional derivative of the fidelity with respect to the condensate
wavefunction $\sigma$, evaluated on a solution of \eqref{EOM1}
\begin{equation}
-\beta_{ij}(g_I,g_J) = \frac{1}{F^2}\left. \frac{\delta^2 F^2 }{\delta \sigma_i(g_I)\delta \sigma_j(g_J)}\right|_{Sol.}
=-\frac{\delta^2 S_{cl} }{\delta \sigma_i(g_I)\delta \sigma_j(g_J)}
\end{equation}
 has to be a \emph{negative definite operator}. Here we are using a compact notation in which the indices
 $i,j=1,2$ refer to the field $\sigma$ ($i,j=1$) or the complex conjugate
 ($i,j=2$), and we are completely suppressing the arguments.

This is the novelty of the approach.
It is a simple test, motivated by an explicit variational method, that allows to estimate
immediately how good is the approximation. The reason for this is simple. The matrix $\beta_{ij}$
can be seen as the second order (functional) derivative of the fidelity at the solution. In particular,
then, it is the lowest order approximation to the peakedness near the maximum, \ie how quickly
the fidelity decrease when the wavefunction is varied. 
It is a rough measure, then of how big is the region
of the space of possible condensate wavefunctions leading to the same qualitatively good approximation as our selected state, in the sense of closeness in the Hilbert space. 
Its eigenvalues, in particular, measure directly the (inverse) of the peakedness of the fidelity
in the direction corresponding to the eigenvector. The larger the eigenvalue, the more
peaked is the fidelity, \ie the better is the approximation.

In particular, it allows us to discard solutions of the classical equations of motion that would
otherwise pass the other self-consistency test (essentially: a lot of small and almost flat tetrahedra, when the fluctuations in the wavefunctions are still large), in cases in which some of the eigenvalues of $\beta_{ij}$ are vanishing or negative. 

The interesting fact is that we can estimate the error in our mean field
approximation \emph{directly}, taking the second functional derivative of the
action and checking whether, on the critical point determined by the equation
of motion, it is a positive definite operator, with its eigenvalues determining the
amplitude of the error made by the simple condensate approximation.

Notice that we have completely bypassed the Schwinger--Dyson hierarchy of equations
for the correlation functions, a fact that can be appreciated even better when considering
dipole states (see Appendix \ref{dipoles}) in comparison with the discussion presented
in \cite{cosmolong}.

In the rest of the paper we will show how this simple idea can be included in a much
more general picture for the derivation of effective equations from GFT.

\section{General case}
We now reconsider the previous result from a general standpoint. The goal is to generalise
the ideas presented in order to extend the class of trial states to mixed states adapted
directly to physical observables (for instance, the average intrinsic curvature, the average extrinsic curvature, etc.).

The general setup of the particular situation at hand is the one of a timeless quantum theory.
The equations
of motion can be cast in the form of (a family of) constraints acting on states
of a given kinematical Hilbert space:
\begin{equation}
 \CC_{\alpha} \ket{\psi} = 0\, .
\end{equation}
Here $\alpha$ is used to label all the equations of motion available. In the case
of \eqref{specialGFT} it would be $\SU^4$, but it could contain additional
labels if matter fields are included, for instance.
If the theory is nontrivial, there will be a subspace of physical states, not
necessarily one dimensional. There, it is necessary to specify a number of
external conditions to identify the state completely, or to work with a specific
statistical ensemble.

For the time being we will consider the special case of a one dimensional space of solutions. The
system is described by a pure state, $\ket{\Phi}$, and by its associated density matrix
$
 \hat{\rho}_{\Phi} = \ket{\Phi}\bra{\Phi}.
$
This is the most complete description of the system, allowing the
computation of the expectation value of any operator, at least in principle. 

In practice, however, we are interested in or we have access
to the description of the behavior of a small set of observables, and we
should therefore formulate a reduced description that keeps track only of
the degrees of freedom relevant for the particular situation considered. 
Furthermore, despite the existence of an exact
expression for $\ket{\Phi}$ (as we will show later), it might be impossible
to extract directly the desired information. A new description has to be
provided. The optimisation procedure 
at the heart of the reduced description will allow the minimisation
of the errors introduced by the reduction itself, as well as the estimation
of their magnitude. This is crucial whenever we want to assess the validity
of the predictions of the effective theory.

The guiding principle proposed here to deal with
these situations is straightforward: replace the state $\hat{\rho}_{\Phi}$
with another (possibly mixed) state $\hat{\rho}_{red}$, associated to a set of operators that
we want to have control of, that best approximate the properties of the original state for the
selected observables, and that
allows us to deduce in a most systematic way the relevant quantities. This state is then
subject to a maximisation procedure that determines the optimal shape that best
approximate (in the sense of the norm) the desired state.

The interpretation of these states within a quantum cosmology scenario, and, in particular,
with the transition to the classical regime (see, for instance,
\cite{Halliwell}) is not clear.
At this stage, they simply encode the maximal ignorance compatible with the available
information about the state. 

The situation considered in \cite{cosmoshort,cosmolong,Gielen} in the case of states
of GFT can be then examined again from a different angle. There, a certain
\emph{ansatz} is proposed to capture the features of the physical state, in such a way
that the representation is as faithful as possible, at least as far as we consider a limited set
of observables. The ideas presented in this section can be used to analyse the same
problem from a different perspective.

The method of projector operators of Nakajima and Zwanzig is
the most appropriate to deal with problems with incomplete information, \ie when the knowledge
of the state of the system is incomplete (see also \cite{openquantumsystems} and \cite{Balian}, for instance). We are then importing this
method, with suitable modifications, into the GFT framework.

\subsection{Basics}

The method consists in the definition of 
a density matrix adapted to the choice of a family of relevant operators $R = \{\hat{A}_{r} \}$.
This is achieved by acting on the original state (described by a (Hermitian) density matrix) 
with a suitable
projection superoperator (an operator mapping operators into other operators):
$
 \hat{\rho}_{R} = \PP_{R} \hat{\rho}_{\Phi}.
$.
The explicit definition of such operator is not very useful, and the interested reader can find
its construction, for instance, in \cite{Balian}. We will
describe the physical ideas behind the construction in a few lines.
Before that, notice that, as a consequence, we can then rewrite the equations of motion for the physical
states in terms of these other
\begin{equation}
 \CC_{\alpha} \hat{\rho}_{\Phi} = \CC_{\alpha} (\PP + \QQ) \hat{\rho}
 =\CC_{\alpha}\hat{\rho}_{R} + \hat{C}_{\alpha}(Irr.) = 0 \label{NZ},
\end{equation}
where $\QQ_{R} = \id-\PP_{R}$.
Given the timeless nature of the system, this is the very simple form
attained by the Nakajima--Zwanzig equation. It is clear that this equation
offers too little information to be useful.
What can be seen is that the effective equation for the relevant
degrees of freedom is influenced by the contribution of the irrelevant
degrees of freedom, as expected, but not more. In particular, there is no
obvious way to estimate the magnitude of the term associated to the irrelevant
part.

Note that the relevant density matrix does not solve, in general, the constraint
equations of the microscopic theory, unless the density matrix itself is built with
operators that commute with the constraints. This might be
important whenever such constraints are interpreted in terms of gauge symmetries.
In that case the method cannot be applied.

The construction of the relevant density matrix (and hence of the superoperator $\PP_{R}$) follows the very general
principle of maximising  the von Neumann entropy of the state when
certain conditions are fixed.
For instance, the state that maximises the von Neumann entropy
at fixed expectation values for a set of Hermitian operators $\Aop_i$ 
is\footnote{Einstein's convention of summation over repeated indices is intended, unless
stated otherwise.}
\begin{equation}
 \hat{\rho}_{red} = \exp\left( -W[\gamma] - \gamma^{i}\Aop_i \right) \label{densitym} ,
\end{equation}
where $\gamma$ are the thermodynamical variables dual to
the operators $\Aop_i$ and $W[\gamma]$ the analogue of the logarithm of the partition function,
necessary to normalise the state:
\begin{equation}
W(\gamma) := \log\Tr{\exp(-\gamma^{i}\Aop_i)}.
\end{equation}
The relation between $\Aop_i$,  their expectation values $\Aexp_i = \langle \Aop_i \rangle$ and $\gamma^j$ is obtained from $\partial \Tr\hrho/\partial \gamma^i =0$:
\begin{equation}
 \frac{\partial W}{\partial \gamma^i} = -\Aexp_i.
\end{equation}
Notice that, in order to be able to determine the $\gamma^i$ given the expectation values, it is necessary that the matrix $w_{ij} = {\partial^2 W}/{\partial \gamma^i \partial \gamma^j}$ is
nonsingular. 

The states \eqref{densitym} are those favoured from a purely information-theoretic point of view:
given the specification of a number of expectation values, these states are maximizing their
von Neumann entropy. Hence, they are the ``most generic'' states compatible with the
available information (specified by the expectation values of the operators
contained in the set R)\footnote{Incidentally, the case
in which the kernel of the constraints has dimension larger than one requires simply
the use of such mixed states. The construction of these states
require the introduction of chemical potentials dual to a set
of operators, commuting with the constraints, which are used to
further specify the states.}.

Therefore, it is natural to use them to replace the physical system, and to find what
are the equations that are implied by the microscopic dynamics, using the same logic
used in statistical mechanics.

\subsection{Effective equations from the fidelity}

The timelessness of the formalism makes the deduction of the equations for the
expectation values somehow different from the procedure normally used in Statistical Mechanics.
Indeed, in \eqref{densitym} there is no trace of the dynamics of the system. 
In Statistical Mechanics, on the contrary, the dynamics of the system is encoded in the time evolution of the expectation values $\Aexp_i$. This time evolution is inferred starting from the underlying quantum
Liouville equation, and projecting it onto the relevant state \eqref{densitym} \cite{openquantumsystems}. 
A different approach is needed to
translate the constraints $\CC_{\alpha}$ into the effective equations linking the
expectation values $\Aexp_i$.

Instead of the standard treatment, one can observe that the quantity:
\begin{equation}
\Prob(\Aexp_i) = \Tr{\hrho_{R}\hrho_{\Phi}} \label{eq:prob}
\end{equation}
naturally generalises the overlap between the trial state and the exact state to the case in which
the first is a density matrix. It is convenient to remind that the fidelity, for general mixed states, is defined as
\begin{equation}
F(\hrho_1,\hrho_2) = F(\hrho_2,\hrho_1) \equiv \Tr{\left( \hrho_1^{1/2}\hrho_2 \hrho_1^{1/2}\right)^{1/2}},
\end{equation}
 and hence, when the
exact state is in fact represented as a density matrix, built out of a manifold
of states solving the equations of motion, this is the expression that should be used. 
However, in the case in which one
of the two states is pure, fot instance $\hrho_{1} = \ket{\psi}\bra{\psi}$, we obtain:
\begin{equation}
\Tr{\hrho_{1}\hrho_{2}} = \bra{\psi} \hrho_2 \ket{\psi} = F^{2}(\hrho_1,\hrho_2),
\end{equation}
where we have used that, for a pure state, $\hrho^{1/2} = \hrho$.

Therefore we define the effective reduced description in terms of the maximisation of
the fidelity, or, in this special case where we assume that the equations of motion have only one
solution, of \eqref{eq:prob}.
This means that, instead of effective equations deduced from the Liouville equation, one
can identify the optimal states by maximising the function $\Prob$ as a function of the
expectation values.

This procedure leads to two conditions. The first is, obviously, the
extremality condition:
\begin{equation}
\frac{\partial \Prob(\Aexp_j)}{\partial \Aexp_i} = 0\, ,
\end{equation}
which leads directly to the relations between the expectation values that are
usually determined as consequences of the equations of motion.
The second is the condition that the matrix
\begin{equation}
G^{ij} = \frac{\partial^2 \Prob}{\partial \Aexp_i \partial \Aexp_j}
\end{equation}
is negative definite. 
Using the density matrix \eqref{densitym}, the extremum is identified by:
\begin{equation}
\frac{\partial \Prob(\Aexp_i)}{\partial \Aexp_i} = 
\frac{\partial \gamma^{j}}{\partial \Aexp_{i}} \left( - \frac{\partial 
W}{\partial \gamma^{j}} \Prob - \Tr{\Aop_j \rho_{R} \rho_{\Phi}} \right)=0
\label{effectiveequationgeneral}
\end{equation}
which can be seen as an implicit equation for 
$
{\Aexp}_{i}
$. This equation shows explicitly how the microscopic dynamics (encoded into the state $\Phi$ solving the
equation of motion)  balanced
with the available information on the state determines the values of the observables in the effective equation.

Notice how the system of equations \eqref{effectiveequationgeneral} is a set of implicit
equations for the multipliers $\gamma^i$, which are the only unknowns of the problem at hand.
Notice also that the formalism can be further generalized considering some of the
$\gamma$s (or, better, $\Aexp_i$) as given from outside. An instance in which this situation
is very natural is when one or more of the relevant observables are used as a reference
frame (for instance a scalar field used as clock in cosmological applications), and we want
to answer the question about the value of the other observables when the value of these
reference variables is specified.

Obviously there can be several solutions to these equations. This situation might
signal that the \emph{ansatz} is degenerate, for this problem: due to symmetries,
for instance, several different density matrices might lead to the same value for the fidelity.
This has to be seen not as a problem, but as a welcome feature of the method. The structure of the
space of solutions to the conditions that arise from the maximisation of the fidelity
might reveal symmetries of the solution that are not evident from the inspection of the
equations of motion.

\subsection{Estimating the accuracy}

The matrix $G^{ij}$ gives a first estimate of the accuracy with which the
physical prediction of the model are given, as it defines the width of the probability
distribution $\Prob(\Aexp_i)$ around the local maxima. In particular, using the suggestive
notation
$
\Prob(\Aexp_i) = \exp(S(\Aexp_i)),
$
the conditions become
\begin{equation}
\frac{\partial S}{\partial \Aexp_i} = 0,
\end{equation}
with the Hessian matrix
\begin{equation}
 \beta^{ij} = -G^{ij} \exp (-S(\overline{\Aexp_i})) \label{eq:betamatrix}
\end{equation}
positive definite.
Around the maxima,
\begin{equation}
\Prob(\overline{\Aexp}_i+\delta\Aexp_i) \approx \Prob(\overline{\Aexp}_i)
\exp\left(-\frac{1}{2}\beta^{ij}\delta \Aexp_i \delta \Aexp_j\right),
\end{equation}
which indicates that $\beta$ controls the region of the space of expectation values for
which the relevant density matrix can be still considered an approximation of the
physical state, albeit not an optimal one. In other words, the peakedness of the function $\Prob$
determines the accuracy with which the physical predictions will be given.

These observations can in turn be related to the typical
fluctuations around the mean value associated to the particular \emph{ansatz}.
The nature of them is unavoidably associated to the reduced representation that we are choosing: if it were possible
to define the notion of equilibrium as usual, these would be the thermodynamic
fluctuations that are associated to our reduced description. 

If we were using $\Prob(\Aexp_i)$ to weight macroscopic states, we could then
estimate the typical fluctuation around a given maximum in the usual way
\begin{align}
\Delta_{Stat} \left( \Aexp_i\Aexp_j \right)_{\bar{\Aexp}}\equiv  \int d\Aexp \Prob(\Aexp)
\langle (\Aop_i-\bar{\Aexp}_i)(\Aop_j-\bar{\Aexp}_j) \rangle_{\Aexp}
\end{align}
where two averages are introduced. The first is the statistical average related to the different
weights provided by
the probability associated to the special choice of the class of quantum state, controlled
by $\Prob$, while the second average is the usual quantum average on the given state, at
fixed value of $\Aexp$, and denoted with $\langle \rangle_{\Aexp}$.

The matrix $\beta$, then, controls the fluctuations about the local maximum value chosen.
Therefore, the matrix $\beta$ measures the reliability of the reduced
density matrix in representing the exact state. Its properties are related to the theoretical error
on the predictions of the effective equations, as it measures the error that is made in
the approximation of the physical state with another state, possibly easier to manipulate.

We now discuss
in more detail the implementation of the method, refining what has been already noticed
in the case of simple condensates.
In particular, for the program to work, one has to be able to have as much information
as possible about $\hrho_{\Phi}$. 

\section{General observables}

We now consider the viability of the
method that we have discussed so far when applied to generic observables.
Consider the case in which the state is taken to be a mixed state which
is optimized for the expectation value of a given Hermitian operator $\OOO[\hphi,\hphid]$, which we assume
to be normal ordered, and which depends on some given kernel that we denote with $\OO$.
The projection on the state, $\Prob$, is now reduced to a functional of the thermodynamic
potential $\gamma$:
\begin{equation}
 \Prob[\gamma]:=\Tr{\rho_{\OO}\rho_{\Phi}} = 
 \frac{Z_0}{Z_{GFT}}\bra{\Phi} \hrho_{\OO} \ket{\Phi}
\end{equation}
Inserting two integrations over field coherent states gives:
\begin{align}
 \Prob[\gamma]& =\frac{Z_{0}}{Z_{GFT}}\frac{1}{Z_0^2}
 \int \DD\psi \DD\overline{\psi} \DD\psi' \DD\overline{\psi'}
 \bra{\psi} \hrho_{\OO}\ket{\psi'}  \times\eol
 \exp\left(
 -\frac{1}{2}\int |\psi|^2 -\lambda V[\psi]
 -\frac{1}{2}\int |\psi'|^2 -\lambda^{*}V[\overline{\psi'}] 
 \right)=\eol
=
 \frac{Z[\OO,\gamma]}{Z_{GFT}Z_0}
 \label{genfun}
\end{align}
In this case, then, the fidelity functional $\Prob$ can be expressed as the
partition function of a certain GFT for \emph{a pair} of complex fields coupled to
the external source. The precise structure of this source term depends on the operator
$\hat{O}$ through the matrix elements
$
 \bra{\psi} \hrho_{\OO}\ket{\psi'} 
$. Since $\hat{O}$ has to contain both $\hphid$ and $\hphi$, the explicit
calculation of the matrix elements involves, in general, nontrivial
manipulations. 
These matrix elements will be determined by the kernel $\OO$ and the conjugate variable
$\gamma$. 

From a field theory point of view, the expression \eqref{genfun} can be seen as
the generating functional for the correlation functions associated to the composite
operator of a certain GFT, as it is determined by the observable $\OOO[\hphi,\hphid]$.
We can then determine the configuration that maximises the fidelity by checking
\begin{equation}
 \frac{\partial \Prob[\OO]}{\partial \gamma} = \frac{1}{Z_0 Z_{GFT}} 
 \frac{\partial Z[\OO,\gamma]}{\partial \gamma}=0\, ,
\end{equation}
which is nothing else than a statement about the vev of this composite operator.
In particular, remembering the structure of $\hrho_{R}$,
\begin{equation}
\hrho_{\OO} = \exp\left( -W(\gamma) - \gamma \OOop[\hphi,\hphid]\right)
\end{equation} where $\gamma$ is a Lagrange multiplier conjugate to the
operator $\OO$, we conclude that:
\begin{equation}
Z[\OO,\gamma] = Z_{aux}[\OO,\gamma] \exp\left(-W(\gamma)\right),
\end{equation}
It is convenient to consider the ``free energy'',
$
\WW=-  \log Z[\OO,\gamma]
$:
\begin{equation}
\frac{\partial \WW}{\partial \gamma} =  \frac{\partial W_{aux}}{\partial \gamma} + \frac{\partial W}{\partial \gamma} = 0
\end{equation}
This is an implicit equation for $\gamma$, and hence for $\langle \OOO \rangle$, that
fixes it in terms of the vacuum expectation value of a certain composite operator
for a certain auxiliary group field theory, that can be computed, in principle.

As a special case, we can consider an optimised state
associated to a field bilinear:
\begin{equation}
\OO[\hphi,\hphid] = \int \OO(g,g')\hphid(g)\hphi(g')\, .
\label{eq:bilinear}
\end{equation}
In this case, we have to consider:
\begin{equation}
 \bra{\psi} \hrho_{\OO}\ket{\psi'} = \exp\left(-W[\gamma,\OO]\right)
 \bra{\psi} \sum_{n=0} \frac{(-\gamma)^n}{n!} \OO[\hphi,\hphid]^n\ket{\psi'}
 \label{matrixelement} 
\end{equation}
The normalisation factor $W[\OO]$ is uniquely determined by the kernel $\OO$ alone.
Using Wick's theorem, we can then massage the RHS of the previous equation and
obtain:
\begin{equation}
 \bra{\psi} \hrho_{\OO}\ket{\psi'} = \exp\left(-W[\gamma,\OO]\right)
 \exp\left( - \Omega[\psi,\psi';\gamma,\OO] \right),
\end{equation}
where $\Omega$ is an implicit expression for the last term in \eqref{matrixelement}.

Therefore, the operator $\OO$ will be controlled by an effective field theory that can
be derived, at least in principle, from the microscopic theory. However, the explicit form of the former might
differ from the latter. This is due to the presence of various corrections due to normal
ordering, essentially, originated from the fact that the operator $\OO$ has to be Hermitian,
in order for $\hrho_{\OO}$ to represent a mixed state, and hence it has to contain simultaneously
creation and annihilation operators. The explicit derivation of the effective field theory, then,
becomes rather complicated.

To give the flavour of the complications involved, we consider the case of operators like \eqref{eq:bilinear}. Indeed, with some manipulations, the action of the exponential of this operator on a coherent state
changes it into another coherent state,
$
\exp(-\gamma \OOO) \ket{\psi} \propto 
\ket{\exp_{\circ}(-\gamma\OO) \psi}
$,
which now depends on the operator $\OOO$. The derivation of the result is given in Appendix \ref{transformation}.
This shows, in fact, that the fidelity will be a functional not of $\OO$, but of its exponential
(as defined in the appendix), which might require then further work to be put into a useful form.

For instance, one could use the crudest approximation and neglect the 
normal ordering corrections:
\begin{align}
& \frac{\bra{\psi} e^{-\gamma\OO}\ket{\psi'}}{ \braket{\psi}{\psi'}} \sim  \frac{\bra{\psi} : e^{-\gamma \OO} :\ket{\psi'}}{\braket{\psi}{\psi'}} \nonumber\\
&= \exp\left( - \int dg_I dg_J \gamma\OO(g_I, g_J)\overline{\psi(g_I)} \psi'(g_J) \right)
 \end{align}
From here, then, one can try to obtain the effective equations for the expectation
value of $\OOO$ and the relative fluctuations along the lines described here.
In particular, one would have to compute the effective action of the field theory
coupled to the two particle source determined by $\OO$, a well defined field theoretic
problem.
If we were interested in the effective equations for simple operators like the total volume,
as implied by the physical state,
one could then obtain them, in principle, with some approximations which now have to
be carefully reconsidered.

Unfortunately, the calculations involved are not easier than those
to be performed if we tried to compute the value of the expectation values of the relevant
composite operators directly  in the exact theory, by computing the associated quantum effective action. 
Therefore, the method presented here has to be supplemented by further
assumptions and simplifications (which require further work for the estimate
of the error), to be successful. For instance, if it were possible
to conclude that in the relevant limit the physical state reduces to a coherent state, this
method will be far more efficient to get an effective equation, rather than the computation
of the corresponding quantum effective action. 

The case of condensed states, then, represents a very special case in which the method based
on the fidelity leads straightforwardly to manageable equations. In fact, one can embed them
in the general formalism that we have considered in these two sections.
They can be seen as a 
case where an infinite number of conditions on the state are imposed. For simple condensates, all the correlation functions are factorized in terms of the one point correlation function 
(\ie all the $n-$point connected correlation functions with $n\geq2$ vanish), for dipoles, all
the odd correlation functions vanish, while the even ones are factorized in terms of the two
point correlation function \cite{cosmolong}, etc.. The state resulting from the imposition
of all these conditions is in fact a pure state,
parametrized by a single wavefunction. This will be then fixed by the variational procedure.
Therefore, this class of trial states, while being only a special case of a much more general framework, and not necessarily adapted to interesting physical observables, represents a privileged starting point for explicit calculations.

\section{Summary}

The elaboration of a set of criteria to be used to decide whether a given \emph{ansatz} for
a state in GFT is a good choice to capture the relevant features the theory 
is a rather pressing issue, when attempting a connection with physical predictions.

In the case of the states envisaged in \cite{cosmoshort,cosmolong}, a number
of self-consistency conditions can be elaborated on the basis of the geometrical picture
that they convey. However, some further conditions
might be obtained only through a careful analysis of the method used to derive the
approximate \emph{ansatz} itself, which might not be visible from the geometrical
point of view.

To address this issue, an information-theoretic path is proposed, following the
approach used in several situations in statistical mechanics, whenever only a limited
amount of information about the microscopic state is available or relevant.
Using an adapted version of the Nakajima--Zwanzig projection method for generic mixed states, it is possible to formulate a systematic derivation of effective equations for reduced density matrices,
even in the case of timeless quantum systems. The key ingredient is the use of the fidelity,
the projection of the trial state onto the physical state.
Effective equations for approximate states for GFT can be put in the form of extremisation
of the fidelity, which in turns leads to the extremisation of
 (auxiliary) effective actions\footnote{Strictly speaking, the various quantities appearing in the preceding sections are at best
only formal expressions, at this stage. It is understood that an appropriate renormalisation
procedure will allow to give proper definitions.} for the relevant variational parameters. 
This method naturally replaces the procedure of minimisation of the energy used to determine the ground state, concepts that are not yet well defined in a timeless formalism.

In the case in which the trial state is modeled after the condensed states
(coherent states, squeezed states etc.), the method leads to explicit equations for the
effective theory in a straightforward way, consistently with previous results and the 
general expectations. Furthermore, a precise estimation of the
theoretical error associated to the effective equations can be given in terms of the
second functional derivative of the fidelity functional.
This allows a first assessment
of the validity of the results, even in terms of just the effective equations themselves, a clear
diagnostic tool that signals whether the results can be reliable or not. This extends
the class of consistency conditions discussed in previous works with a criterion
which is essentially the Ginzburg criterion for the validity of the mean field approximation.

In the case of a simple condensate, \ie a coherent state,
the method proposed here shows that the best estimate for the equations of motion to be used is given by the expression \eqref{EOM1} involving \emph{only} the classical action. At the same time, the reasoning that
we have proposed implies naturally that the error arising from this mean field
approximation is indeed encoded into the second functional derivative of the same classical action, which is straightforward
to compute.
The  eigenvalues of the resulting differential operator reflect the
fluctuation properties around the ansatz, \ie the validity of the approximation, as they are
inversely proportional to the peakedness of the fidelity functional around the maximum.

The same principles can be applied to the case of dipole states (see Appendix \ref{dipoles}).
Despite the fact that an explicit functional cannot be given, an order by
order expansion of the fidelity in the dipole wavefunction 
leads to an effective field
equation which now includes the effect of the interaction term.
This is an improvement with respect to the situation considered in \cite{cosmolong}, where
it was shown that, following the Schwinger--Dyson hierarchy, the interaction term and the kinetic
term for a Boulatov--Ooguri model would appear in distinct equations. This separation
might be unpleasant, as the interaction term includes some of the details of the implementation
of the simplicity constraint (technically, how the representations of $\SU$ are embedded in those of $Spin(4)$ for Riemannian and $\mathrm{SL}(2,\mathbb{C})$ for Lorentzian gravity), turning an otherwise BF topological amplitudes into gravitational amplitudes. Furthermore, it contains the combinatorial
structure that leads to the interpretation of transition amplitudes in terms of four dimensional simplicial complexes.

It is worth noting that a failure of the concavity criterion can still
provide useful information, even though it would lead to the
rejection of the given solution.  
This is the case of the linear equation
considered in \cite{cosmoshort,cosmolong,Gielen,GielenOriti,Calcagni} as a toy model.
There, the presence of a nontrivial kernel for the linear operator acting on the wavefunction shows
that additional conditions need to be specified to pinpoint a solution.
In general, the appearance of more than one solution of the Nakajima--Zwanzig
equations requires the specification of additional quantities, in the trial states,
that can be controlled externally and that will influence the resulting dynamics.

The presence of modes with positive eigenvalues will signal instabilities, \ie a saddle point
of the action, and is related to an incorrect or a degenerate guess.  
This might signal, for instance, that the state obtained by solving the quantum equations of
motion in perturbation theory is not the only physical state.
The presence of zero modes, instead, might be more intriguing as it might signal the
presence of additional symmetries (generating flat directions) 
that have not been correctly included.

These considerations match and extend those mentioned in \cite{cosmolong}, 
encapsulating them in a more general framework that allows to consider, at least in principle, a wider
set of observables. For instance, it might be possible, with suitable approximations, to adapt the formalism to deal immediately with
one-particle operators like the total volume, bypassing completely the 
definition of a ``condensate wavefunction'', the solution of equations of motion and the
computation of the desired expectation value on the resulting state. 
This is particularly urgent as it is clear that, at least in the case of simple condensates,
the superspace over which the condensate wavefunction is defined is the configuration
space of the quantum tetrahedron \cite{Barbieri}, which differs rather dramatically from the configuration
space of the classical tetrahedron, expected and needed if a semiclassical analysis
is considered. However, at this stage it is unclear whether it is possible to organize concrete
calculations.

\section{Outlook}

The derivation of the effective equations for GFT condensates (and the associated error) from the fidelity is an interesting subject in itself. However, the generic formalism that has been described
allows for some developments that deserve some further discussion.

The inclusion of fluctuations around the mean field is of obvious interest for the development
of a theory of cosmological perturbations from GFTs and LQG. The framework considered
in this paper might open a road for this sort of calculations.
In this respect, the fidelity (actually, its second functional derivative at the maximum) controls
the fluctuations around what can be interpreted as the ``equilibrium state" in this timeless framework. As the equilibrium state can be interpreted, under certain conditions, as a homogeneous cosmology, the fluctuations encoded in the fidelity will be naturally
associated to deviations from homogeneity. This is a further step in the direction of turning some ideas that were discussed in \cite{cosmolong} into quantitative statements.

While a derivation of a local field theory is still missing,
we can then say that the natural class of effective field theories that we should expect 
to describe the continuum limit of the theory is of the stochastic gravity type \cite{stochastic}, \ie
field theories including stochastic fluctuations of the geometry, with a noise term that describes the random deviations from the equilibrium/maximum fidelity configuration. The noise term entering the Langevin
equation will have to match the statistical fluctuations determined by $\beta_{ij}$, that
can be computed (in principle) in terms of the fidelity as in \eqref{eq:betamatrix}. 

For instance, if we consider the case of simple condensates, we should
expect the effective equation to be not a Gross--Pitaevski equation, but rather a
\emph{stochastic Gross--Pitaevski} one (see, for instance, \cite{stochasticGPE} and references therein), whose noise kernel is determined entirely
by $\delta^2S/\delta\sigma\delta \sigma$.
The analysis of the content of such an equation (and the implications for the inhomogeneities) is left for future work. What it is important to remark is that this is something that now we can compute.
The next
step, then, is to turn the fluctuations around condensates into ordinary metric (and connection)
perturbations.

Second, one might be discouraged from the fact that,
for generic observables, the derivation of the effective theory might be as difficult
as the computations for the exact theory. The reason for this is that a form of coarse graining (leading to possibly very complicated
effective actions) is unavoidable.
This is not unexpected. Indeed, the formulation of kinetic theory
or fluid dynamics does not solve the problem of computing the physical properties of the
fluid itself, once the microscopic Hamiltonian of the system is given.
Nonetheless, the formulation of the problem in the language of
field theory gives the opportunity to use a large number of approximation methods
to be used for concrete calculations. 
It is necessary to introduce approximations and further insights coming from a different
analysis. In fact, we have never used one important point, that might simplify the calculations.

So far we have never used the fact that we are interested not in a generic
regime of the theory, but rather in what will be the continuum limit, which might be associated
to a phase transition of the underlying GFT. Our manipulations have just turned the problem
of investigating the properties of a state into the generation of effective equations, without
really addressing the problem of their solution, or the particular regime in which they should
be considered. It is at this point that some physical insight
has to rescue us, with the development of suitable phenomenological models.

Concretely, one might use the mathematical properties of this phase transition to further simplify the effective equations \emph{directly at the effective level}. For example, if some scaling
regime appears, the equations can be expected to become considerably easier to deal with
\cite{gftemergent}. Another possible scheme is analogous to the high temperature expansion,
\ie the regime in which $\gamma^i$ are so small that the product with the eigenvalues
of the the relevant operators is always much smaller than the unity. This might be an approximation relevant for the description
of the region near the phase transition (if it exists) between the ``no space vacuum'',
associated to the Fock vacuum, and a geometric phase, when the vacuum state shows
the first germs of a macroscopic geometry. 
The implementation of these ideas and the development of the
corresponding approximation schemes is left for future work.

\acknowledgments{I would like to thank F. Caravelli, S. Gielen and D. Oriti for stimulating discussions
on the topic and for constructive comments on a previous version of this work.
This work has been supported by the Templeton Foundation
through the grant number PS-GRAV/1401.}

\appendix

\section{Effective equations for dipoles}\label{dipoles}

For completeness, we consider the case of dipoles. In \cite{cosmoshort,cosmolong},
the derivation of the equations of motion was done in terms of the examination of the tower
of Schwinger--Dyson equations. There are two problems hidden in such an approach.
First, there is no clear prescription on how many equations one should retain. Second,
as it has been shown, the simplest equations, those admitting the interpretation in terms
of a modified Friedmann equation, are built out of the part of the equation of motion that
 \emph{does not} include the details of the operators used to impose the simplicity constraints,
 which is, in fact, the essence of the dynamics of gravity \emph{\'a la} Plebanski. This
 should raise suspicions about the result.

With the fidelity method it is possible to derive effective equations that are retaining all
the dynamical information, as we shall show.

The state under consideration is
\begin{align}
\ket{\xi} = \exp(\hxi)\fv,
\hxi \equiv \frac{1}{2}\int (dg)^8 \xi(g_{I},g_{J})\hphid(g_i)\hphid(g_J) 
\end{align}
apart from a normalisation, which can be computed as a power series in the operator
$\zeta(g_I,g_J) = \int dg_K \xi^{*}(g_I,g_K) \xi(g_K,g_J)$.

The computation of the fidelity
\begin{equation}
\Prob[\xi] = \frac{|\braket{\xi}{\Phi}|^2}{\braket{\xi}{\xi}}
\label{fidelity}
\end{equation}
 for generic $\xi$ is too complicated to be useful. Instead,
we will work in an expansion in $\xi$, which is done for illustrative purposes only.
Indeed, keeping $\xi$ small is equivalent to restrict the analysis to a regime in which
there are very few quanta of the field. Therefore, it is hardly useful for a regime described by
a macroscopic geometry  \cite{cosmolong,Gielen,GielenOriti}.

Due to the quintic nature of the interaction term, the first nontrivial
contribution to the numerator of the fidelity comes from:
\begin{align}
\braket{\xi}{\Phi} = \fvbra \left( \sum_{n=0}^{\infty}
\frac{1}{n!} (\hxid)^n \right)\exp(- \lambda \VV[\hphid])\fv
 \nonumber\\
\sim \fvbra\left(1 - \frac{1}{5!} (\hxid)^5 \right) 
\left( 1 + \frac{1}{2} \lambda^2  \VV^2[\hphid] \right)\fv
\label{numerator}
\end{align} 
This term contains the contraction of five kernels $\xi^{*}$ with 
two kernels $\KV$, which encode the microscopic models,
in all the possible ways as dictated by Wick's theorem.
The next contribution will be then of order $\xi^{10}$, and we neglect it.
Instead of writing explicitly all the terms that we obtain in
the truncation of the numerator, we collectively denote the resulting kernel
as $\KV_{\eff}$. Therefore:
\begin{align}
\braket{\xi}{\Phi} \sim 1 - \frac{\lambda^2}{240} \int (dg)^{40}
\KV_{\eff}(g_{I_{1}},g_{J_{1}},\ldots,g_{I_{5}},g_{J_{5}})
\times \nonumber \\
\xi^{*}(g_{I_{1}},g_{J_{1}})\cdots \xi^{*}(g_{I_{5}},g_{J_{5}})  
\end{align}
The norm of the trial state can be evaluated up to terms of order $\xi^{6}$:
\begin{equation}
\braket{\xi}{\xi} \sim 1 + \frac{1}{2} \Tr{\zeta} + \left( \Tr{\zeta} \right)^2 
+2 \Tr{\zeta^2} + O(\xi^6),
\end{equation}
where $\Tr{\zeta} = \int dg_I \zeta(g_I,g_I)$.
Taking the functional derivative of \eqref{fidelity} with respect to $\xi^*$ (the fact that
the fidelity is real implies that we need just this variation) we obtain the effective equation:
\begin{align}
0 = -\frac{1}{2} \xi(g_{I},g_J)
- \frac{3}{2} \Tr{\zeta} \xi(g_{I},g_J)+\nonumber \\
- \int dg_K dg_{H} \xi(g_{I},g_K)\xi^*(g_{K},g_H)\xi(g_{H},g_J)
+ \nonumber \\
-\frac{\lambda^2}{48}\int (dg)^{32} \KV_{\eff}(g_{I},g_{J},\ldots,g_{I_{5}},g_{J_{5}})
\times \nonumber \\
\xi^{*}(g_{I_{2}},g_{J_{2}})\cdots \xi^{*}(g_{I_{5}},g_{J_{5}}) + O(\xi^5).
\label{effdipole}
\end{align}
This equation has two important properties. It is nonlinear, which means that
the number of quanta, related to the wavefunction $\xi$, is not completely
arbitrary as a solution cannot be rescaled at will. Second, it includes the
first effects of the interaction term, in the same equation.

The second variational derivative could be computed along the same lines. By inspection
of \eqref{effdipole}, we conclude that it will depend nontrivially on the wavefunction $\xi$ solving
that effective equation, and hence it will generally lead to some nontrivial conditions. Its spectrum could be then used to check the viability of $\xi$
as a meaningful solution.

\section{A useful transformation} \label{transformation}
In this appendix we are going to prove that:
\begin{equation}
\exp(+\OOO) \hphi(g_I) \exp(-\OOO) =
\int dg_J\exp(\OO)(g_I,g_J) \hphi(g_J)
\end{equation}
First, notice that:
\begin{equation}
\exp(+\OOO) \hphi(g_I) \exp(-\OOO) = \hphi(g_I)+\sum_{n=1}^{\infty} \frac{1}{n!}[\OOO,\hphi(g_I)]_n
\end{equation}
where
\begin{align}
&[\OOO,\hphi(g_I)]_1 = [\OOO,\hphi(g_I)] \\
&[\OOO,\hphi(g_I)]_n = [\OOO,[\OOO,\hphi(g_I)]_{n-1}], \qquad n\geq 2 
\end{align}
Then, using the commutation relations for the field operators:
\begin{equation}
[\OOO,\hphi(g_I)] = -\int dg_J \OO(g_I,g_J)\hphi(g_J),
\end{equation}
it is possible to show, by induction, that
\begin{equation}
[\OOO,\hphi(g_I)]_n = (-1)^{n} \int dg_J \OO_{n}(g_I,g_J) \hphi(g_J), n\geq 2
\end{equation}
where:
\begin{equation}
\OO_{n}(g_I,g_J) = \int dg_K \OO(g_I,g_K ) \OO_{n-1}(g_K,g_J)
\end{equation}
is the n-th convolution of the kernel $\OO$ with itself.
If we use the shorthand notation:
\begin{equation}
\exp_{\circ}(-\OO)(g_I,g_J) = \delta(g_I,g_J) + \sum_{n=1}^{\infty}
\frac{(-1)^n}{n!} \OO_{n}(g_{I},g_J)
\end{equation}
the final result is:
\begin{equation}
\exp(+\OOO) \hphi(g_I) \exp(-\OOO) =
\int dg_{J}
\exp_{\circ}(-\OO)(g_I,g_J) \hphi(g_J)
\end{equation}
As a consequence, then:
\begin{align}
\hphi(g_I)\exp(-\OOO) \ket{\psi} =\nonumber \\
\exp(-\OOO)\exp(\OOO)
\hphi(g_I)\exp(-\OOO) \ket{\psi} = 
\nonumber \\
= 
\exp(-\OOO)
\int dg_{J}
\exp_{\circ}(-\OO)(g_I,g_J) \hphi(g_J)\nonumber\\
=
\int dg_{J}\exp_{\circ}(-\OO)(g_I,g_J) \psi(g_J) \exp(-\OOO) \ket{\psi} 
\end{align}
This means that:
\begin{equation}
\exp(-\OOO) \ket{\psi} \propto 
\ket{\exp_{\circ}(-\OO) \psi}
\end{equation}
The proportionality factor requires further analysis and is not included here.


\begin{thebibliography}{10}
 \bibitem{cosmoshort}
  S.~Gielen, D.~Oriti and L.~Sindoni,
  Phys.\ Rev.\ Lett.\  {\bf 111} (2013) 031301
  [arXiv:1303.3576 [gr-qc]].
 
\bibitem{cosmolong}
  S.~Gielen, D.~Oriti and L.~Sindoni,
  arXiv:1311.1238 [gr-qc].
  
\bibitem{Gielen}
S.~Gielen,
  Class.\ Quant.\ Grav.\  {\bf 31} (2014) 155009
  [arXiv:1404.2944 [gr-qc]].
  
\bibitem{GielenOriti}
  S.~Gielen and D.~Oriti,
  arXiv:1407.8167 [gr-qc].  
  
\bibitem{Calcagni}
  G.~Calcagni,
  arXiv:1407.8166 [gr-qc].  


\bibitem{Burgess}
  C.~P.~Burgess,
  Living Rev.\ Rel.\  {\bf 7} (2004) 5
  [gr-qc/0311082].

\bibitem{fidelity} P.~Zanardi, N.~Paunkovi\'c, 
Phys.\ Rev. \ {\bf E, 74} (2006) 031123 [arXiv:quant-ph/0512249]

\bibitem{openquantumsystems}
  H.~P.~Breuer and F.~Petruccione,
  ``The theory of open quantum systems,''
  Oxford, UK: Univ. Pr. (2002) 625 p

\bibitem{Balian}
R.~Balian, 
American 
Journal of Physics, {\bf 67} (1999) 1078 

  \bibitem{matrixmodels}
  P.~Di Francesco, P.~H.~Ginsparg and J.~Zinn-Justin,
  Phys.\ Rept.\  {\bf 254} (1995) 1
  [hep-th/9306153].

  
    \bibitem{GurauRyan}
  R.~Gurau and J.~P.~Ryan,
  SIGMA {\bf 8} (2012) 020
  [arXiv:1109.4812 [hep-th]].



\bibitem{OritiMicroDyn}
  D.~Oriti,
  arXiv:1110.5606 [hep-th].  
  
  
  \bibitem{GFTnoncomm}
  A.~Baratin and D.~Oriti,
  Phys.\ Rev.\ Lett.\  {\bf 105} (2010) 221302
  [arXiv:1002.4723 [hep-th]].  
  
\bibitem{Baratin}
  A.~Baratin, B.~Dittrich, D.~Oriti and J.~Tambornino,
  Class.\ Quant.\ Grav.\  {\bf 28} (2011) 175011
  [arXiv:1004.3450 [hep-th]].    


  \bibitem{2nd} 
  D.~Oriti,
  arXiv:1310.7786 [gr-qc].
  
\bibitem{Ooguri}
  H.~Ooguri,
  Mod.\ Phys.\ Lett.\ A {\bf 7} (1992) 2799
  [hep-th/9205090].



\bibitem{Perez}
  A.~Perez,
  Living Rev.\ Rel.\  {\bf 16} (2013) 3
  [arXiv:1205.2019 [gr-qc]].



\bibitem{Barbieri}
  A.~Barbieri,
  Nucl.\ Phys.\ B {\bf 518} (1998) 714
  [gr-qc/9707010].


\bibitem{Halliwell}
  J.~J.~Halliwell,
  Phys.\ Rev.\ D {\bf 39} (1989) 2912.



%
\bibitem{stochastic}
  B.~L.~Hu and E.~Verdaguer,
  Living Rev.\ Rel.\  {\bf 11} (2008) 3
  [arXiv:0802.0658 [gr-qc]].

\bibitem{stochasticGPE}
 E.~Calzetta, B.~L.~Hu and E.~Verdaguer,
 Int.\ J.\ Mod.\ Phys.\ {\bf B 21} (2007) 4239 [cond-mat/0702046]


\bibitem{gftemergent}
  L.~Sindoni,
  arXiv:1105.5687 [gr-qc].



  
  
  




\end{thebibliography}
\end{document}